\documentclass[9pt,twocolumn,twoside,]{pinp}

\providecommand{\tightlist}{%
  \setlength{\itemsep}{0pt}\setlength{\parskip}{0pt}}

\usepackage[T1]{fontenc}
\usepackage[utf8]{inputenc}

\definecolor{pinpblue}{HTML}{185FAF}  % imagecolorpicker on blue for new R logo
\definecolor{pnasbluetext}{RGB}{101,0,0} %

\title{An R Autograder for PrairieLearn}

\author[1]{Dirk Eddelbuettel}
\author[1]{Alton Barbehenn}

  \affil[1]{Department of Statistics, University of Illinois, Urbana-Champaign, IL,
USA}

\leadauthor{Eddelbuettel + Barbehenn}

\begin{abstract}
We desribe how we both use and extend the PrarieLearn framework by
taking advantage of its built-in support for \emph{external}
auto-graders. By using a custom Docker container, we can match our
course requirements perfectly. Moreover, by relying on the flexibility
of the interface we can customize our Docker container. A specific
extension for unit testing is described which creates context-dependent
difference between student answers and reference solution providing a
more comprehensive response at test time.
\end{abstract}

\dates{This version was compiled on \today} 

\doifooter{\url{https://github.com/stat430dspm/rocker-pl}}
\pinpfootercontents{R Autograder for PrairieLearn}

\begin{document}

% Optional adjustment to line up main text (after abstract) of first page with line numbers, when using both lineno and twocolumn options.
% You should only change this length when you've finalised the article contents.
\verticaladjustment{-2pt}

\maketitle
\thispagestyle{firststyle}
\ifthenelse{\boolean{shortarticle}}{\ifthenelse{\boolean{singlecolumn}}{\abscontentformatted}{\abscontent}}{}

% If your first paragraph (i.e. with the \dropcap) contains a list environment (quote, quotation, theorem, definition, enumerate, itemize...), the line after the list may have some extra indentation. If this is the case, add \parshape=0 to the end of the list environment.

\hypertarget{context}{%
\section{Context}\label{context}}

We describe the motivation, design and use of an autograder for the R
language within the PrairieLearn system \citep{talk:pl:2018}.
PrairieLearn is in use at the University of Illinois at
Urbana-Champaign, where it is also being developed, and other campuses
to support fully automated computer-based testing of homework, quizzes
and examples for undergraduate and graduate students. We use it to
support the topics course \href{https://stat430.com}{STAT 430 ``Data
Science Programming Methods''} we have been teaching since 2019 in the
Department of Statistics at the University of Illinois at
Urbana-Champaign.

As documented, PrairieLearn supports \emph{external graders}, and we are
providing one such grader for the R language and system. Our
implementation follows
\href{https://en.wikipedia.org/wiki/KISS_principle}{KISS principles},
and is sufficiently safe and robust for deployment. Our approach uses
two key insights. First, testing student submissions is close to unit
testing code---and we benefit from relying on a very clever, small and
nimble test framework package, \textbf{tinytest} \citep{CRAN:tinytest}.
Second, the PrairieLearn decision to allow external graders under a
`bring your own container' scheme allows us to regroup all our
requirement in a simple Docker container---extending a base container
from the Rocker Project \citep{rocker:2017}---we provision and control.

\hypertarget{prairielearn}{%
\section{PrairieLearn}\label{prairielearn}}

PrairieLearn \citep{talk:pl:2018} is an online problem-driven learning
system for creating homeworks and tests that enables automated code
evaluation as well as more traditional question types (like multiple
choice questions) for both homework assignments as well as exams. It is
built to be flexible, and enables grading to happen however the
instructor wishes using Docker. PrairieLearn comes with many easy ways
of adding randomization to questions, and a custom set of HTML tags that
makes writing questions easy.

\hypertarget{direct-prairielearn-integration}{%
\section{Direct PrairieLearn
Integration}\label{direct-prairielearn-integration}}

The integration between different components is designed to be simple
and flexible. Data is exchanged by configuration text files in the JSON
format (which is discussed below). At its core, this involves only two
files (which we describe next) that are made available in the top-level
directory of the contributed grader as shown the following listing:

\begin{Shaded}
\begin{Highlighting}[]
\NormalTok{fs}\OperatorTok{::}\KeywordTok{dir_tree}\NormalTok{(}\StringTok{"r_autograder"}\NormalTok{)}
\CommentTok{#  r_autograder}
\CommentTok{#  +-- pltest.R}
\CommentTok{#  \textbackslash{}-- run.sh}
\end{Highlighting}
\end{Shaded}

\hypertarget{run.sh}{%
\subsection{run.sh}\label{run.sh}}

The first file, \texttt{run.sh}, shown in Appendix 1, is more-or-less
unchanged from the \texttt{run.sh} file in the PrairieLearn example
course which invokes the file \texttt{pltest.R} discussed next. It sets
up a number of environment variables reflecting the PrairieLearn setup.
It also copies files in place, adjusts modes (more on that below when we
discuss security), calls the evaluation script discussed next, and
assembles the result.

\hypertarget{pltest.r}{%
\subsection{pltest.R}\label{pltest.r}}

The second file is the actual test runner for PrairieLearn under R, and
is shown in Appendix 2.

For a given question, essentially three things happen:

\hypertarget{extract-metadata-from-tests}{%
\subsubsection{Extract Metadata from
Tests}\label{extract-metadata-from-tests}}

The displayed title of each available test, and the available points per
test, are extracted from the question's test files themselves. This is
performed by the helper function \texttt{plr::get\_question\_details()}
which is given the default test directory used in our layout:
\texttt{"/grade/tests/tests"}.\footnote{Earlier or alternate approaches use
an explicit file \texttt{points.json}; we find it more suitable to define this on
the fly given the test files.} Our implementation is inspired by the
doxygen and roxygen2 tag use for documentation and is most appropriate:
metadata for each tests is stored with test. This allows for quick
iteration during development as test files can simply be renamed to be
deactivated without worrying about conflicting metadata.

\hypertarget{run-the-tests}{%
\subsubsection{Run the Tests}\label{run-the-tests}}

The actual test predicates are being run using the \textbf{tinytest}
package and its function \texttt{run\_test\_dir()} traversing a
directory (more on that below). The result is then converted into a
data.frame object. We discuss the \textbf{tinytest} framework in more
detail below.

\hypertarget{merge-and-post-process}{%
\subsubsection{Merge and Post-Process}\label{merge-and-post-process}}

The two data.frame objects (metadata and test results) are merged using
the names of each test file as the key. Then points are calculated and
the resulting object is written as a JSON file for PrairieLearn to
consume.

\hypertarget{test-framework}{%
\section{Test Framework}\label{test-framework}}

\textbf{tinytest} is an appropriately light-weight test framework
without further dependencies. As stated in the opening of its vignette:

\begin{quote}
The purpose of \emph{unit testing} is to check whether a function gives
the output you expect, when it is provided with certain input.
\end{quote}

This is precisely what checking student answers amounts to. Given the
context of a question, students provide code, typically as a function,
which we can test given inputs---and compare to a reference answer and
its output. Our framework does just that.

Two of the key insights of \textbf{tinytest} are:

\begin{enumerate}
\def\labelenumi{\alph{enumi})}
\tightlist
\item
  test results are data which can be stored and manipulated, and
\item
  that each test file is a \emph{script} interspersed with command and
  suitable to be programmed over.
\end{enumerate}

We use another key feature of \textbf{tinytest}: its extensibility. Our
small helper package \textbf{ttdo} \citep{CRAN:ttdo} extends the
\textbf{tinytest} framework by using \textbf{diffobj}
\citep{CRAN:diffobj} to compute succinct \texttt{diff(1)}-style
summaries of object comparisons. This is most useful to show students
the differences between their result and the reference result. We show
this below in the context of a question.

\hypertarget{example-r-question}{%
\section{Example R Question}\label{example-r-question}}

Within the testing framework, questions are a key component. In general,
each question is organized in its own directory. Questions may then be
grouped by directory name for assignments, exams or quizzes comprising a
set of such questions.

For each question used in our autograder, the directory layout is as
shown in the next figure.

\begin{Shaded}
\begin{Highlighting}[]
\NormalTok{fs}\OperatorTok{::}\KeywordTok{dir_tree}\NormalTok{(}\StringTok{"rfunction-fib"}\NormalTok{)}
\CommentTok{#  rfunction-fib}
\CommentTok{#  +-- info.json}
\CommentTok{#  +-- initial_code.R}
\CommentTok{#  +-- question.html}
\CommentTok{#  \textbackslash{}-- tests}
\CommentTok{#      +-- ans.R}
\CommentTok{#      \textbackslash{}-- tests}
\CommentTok{#          +-- test_00_fib1.R}
\CommentTok{#          +-- test_01_fib2.R}
\CommentTok{#          \textbackslash{}-- test_02_fibn.R}
\end{Highlighting}
\end{Shaded}

There are two mandatory top-level files:

First, \texttt{info.json} which contains all the relevant data for this
question, including of course which grader to use. As discussed above,
this file controls \emph{which} of several graders is used.

\begin{Shaded}
\begin{Highlighting}[]
\FunctionTok{\{}
  \DataTypeTok{"uuid"}\FunctionTok{:} \StringTok{"32A98E04-0A4C-497A-91D2-18BC4FE98047"}\FunctionTok{,}
  \DataTypeTok{"title"}\FunctionTok{:} \StringTok{"Fibonacci Sequence 2.0"}\FunctionTok{,}
  \DataTypeTok{"topic"}\FunctionTok{:} \StringTok{"Functions"}\FunctionTok{,}
  \DataTypeTok{"tags"}\FunctionTok{:} \OtherTok{[}\StringTok{"code"}\OtherTok{,} \StringTok{"v3"}\OtherTok{,} \StringTok{"barbehe2"}\OtherTok{,} \StringTok{"deddel"}\OtherTok{,} 
           \StringTok{"balamut2"}\OtherTok{,} \StringTok{"stat430dspm"}\OtherTok{,} \StringTok{"Fa19"}\OtherTok{,}
           \StringTok{"rautograder"}\OtherTok{]}\FunctionTok{,}
  \DataTypeTok{"type"}\FunctionTok{:} \StringTok{"v3"}\FunctionTok{,}
  \DataTypeTok{"singleVariant"}\FunctionTok{:} \KeywordTok{true}\FunctionTok{,}
  \DataTypeTok{"gradingMethod"}\FunctionTok{:} \StringTok{"External"}\FunctionTok{,}
  \DataTypeTok{"externalGradingOptions"}\FunctionTok{:} \FunctionTok{\{}
    \DataTypeTok{"enabled"}\FunctionTok{:} \KeywordTok{true}\FunctionTok{,}
    \DataTypeTok{"image"}\FunctionTok{:} \StringTok{"stat430/pl"}\FunctionTok{,}
    \DataTypeTok{"serverFilesCourse"}\FunctionTok{:} \OtherTok{[}\StringTok{"r_autograder/"}\OtherTok{]}\FunctionTok{,}
    \DataTypeTok{"entrypoint"}\FunctionTok{:} \StringTok{"/grade/server/r_grader/run.sh"}\FunctionTok{,}
    \DataTypeTok{"timeout"}\FunctionTok{:} \DecValTok{5}
  \FunctionTok{\}}
\FunctionTok{\}}
\end{Highlighting}
\end{Shaded}

We note that this points specifically to

\begin{itemize}
\tightlist
\item
  a top-level directory (such as the one shown above),
\item
  an entry-point script (as discussed above)
\item
  a container to run the evaluations in.
\end{itemize}

Second, \texttt{question.html} which defines the display shown to the
student. PrairieLearn now allows for markdown to describe the central
part, and can reference external files such as the file
\texttt{initial\_code.R} listed here too. \texttt{initial\_code.R}
provides the stanza of code shown in the Ace editor component (and the
file name is specified in \texttt{question.html}).

Then, the \texttt{tests/} directory contains the test infrastructure. By
our convention, \texttt{tests/ans.R} is the reference answer. This file
is set to mode 0600 to ensure the student code can never read it.

\begin{Shaded}
\begin{Highlighting}[]
\CommentTok{# Reference answer to find nth term in the }
\CommentTok{# Fibonacci sequence using non-recursive methods}
\NormalTok{fib <-}\StringTok{ }\ControlFlowTok{function}\NormalTok{(n) \{}
\NormalTok{  out <-}\StringTok{ }\KeywordTok{rep}\NormalTok{(}\DecValTok{1}\NormalTok{, n)}
  \ControlFlowTok{if}\NormalTok{ (n }\OperatorTok{>=}\StringTok{ }\DecValTok{3}\NormalTok{) }
     \ControlFlowTok{for}\NormalTok{ (i }\ControlFlowTok{in} \DecValTok{3}\OperatorTok{:}\NormalTok{n)}
\NormalTok{       out[i] <-}\StringTok{ }\NormalTok{out[i}\DecValTok{-1}\NormalTok{] }\OperatorTok{+}\StringTok{ }\NormalTok{out[i}\DecValTok{-2}\NormalTok{]}
  \KeywordTok{return}\NormalTok{(out)}
\NormalTok{\}}
\end{Highlighting}
\end{Shaded}

The subdirectory \texttt{tests/tests/} then contains one or more
\emph{unit tests} or, in our case, question validations. The first
question sources the file, evaluates \(F(1)\) and compares to the
expected answer, 1. (Other test questions then check for other values as
shown below; several test predicates could also be present in a single
test file but we are keeping it simple here.)

\begin{Shaded}
\begin{Highlighting}[]
\CommentTok{## @title Test F(1)}
\CommentTok{## @score 2}

\NormalTok{file <-}\StringTok{ "/grade/student/fib.R"}
\NormalTok{v <-}\StringTok{ }\NormalTok{plr}\OperatorTok{::}\KeywordTok{source_and_eval_safe}\NormalTok{(file, }\KeywordTok{fib}\NormalTok{(}\DecValTok{1}\NormalTok{), }\StringTok{"ag"}\NormalTok{)}

\KeywordTok{expect_equal}\NormalTok{(v, }\DecValTok{1}\NormalTok{)}
\end{Highlighting}
\end{Shaded}

Of note is our use of a function from the helper package \textbf{plr}
\citep{R:plr}. As the same code fragment would be repeated across
numerous question files, it makes sense to regroup this code in a
(simple) function. At its core are the \texttt{system()} call, made as
the autograde user \texttt{ag}, and the subsequent evaluation of the
supplied expression. We take full advantage of the lazy evaluation that
makes R so powerful: \texttt{fib(1)} is not evaluated by the caller but
rather in the context of the caller---after sourcing the corresponding
file. We also make the file to sourced visible to the \texttt{ag} user.
All other files remain inaccessible thanks for their mode of 0600.

Another key aspect is the use of \texttt{eval\_safe()} from the
\textbf{unix} package \citep{CRAN:unix}. As we are effectively running
as \texttt{root} inside a container, we have the ability to lower to
permission to those of another user, here \texttt{ag}.

\begin{Shaded}
\begin{Highlighting}[]
\CommentTok{## roxygen2 documentation omitted here, see repo}
\NormalTok{source_and_eval_safe <-}\StringTok{ }\ControlFlowTok{function}\NormalTok{(file, expr,}
                                 \DataTypeTok{uid=}\OtherTok{NULL}\NormalTok{) \{}
  \ControlFlowTok{if}\NormalTok{ (}\OperatorTok{!}\KeywordTok{is.null}\NormalTok{(uid) }\OperatorTok{&&}
\StringTok{      }\KeywordTok{class}\NormalTok{(uid) }\OperatorTok{==}\StringTok{ "character"}\NormalTok{)}
\NormalTok{    uid <-}\StringTok{ }\NormalTok{unix}\OperatorTok{::}\KeywordTok{user_info}\NormalTok{(uid)}\OperatorTok{$}\NormalTok{uid}

  \ControlFlowTok{if}\NormalTok{ (}\OperatorTok{!}\KeywordTok{file.exists}\NormalTok{(file)) }\KeywordTok{return}\NormalTok{(}\KeywordTok{invisible}\NormalTok{(}\OtherTok{NULL}\NormalTok{))}

\NormalTok{  oldmode <-}\StringTok{ }\KeywordTok{file.mode}\NormalTok{(file)}
  \KeywordTok{Sys.chmod}\NormalTok{(file, }\DataTypeTok{mode=}\StringTok{"0664"}\NormalTok{)}
  \KeywordTok{source}\NormalTok{(file)}

\NormalTok{  res <-}\StringTok{ }\NormalTok{unix}\OperatorTok{::}\KeywordTok{eval_safe}\NormalTok{(expr, }\DataTypeTok{uid=}\NormalTok{uid)}
  \KeywordTok{Sys.chmod}\NormalTok{(file, }\DataTypeTok{mode=}\NormalTok{oldmode)}
\NormalTok{  res}
\NormalTok{\}}
\end{Highlighting}
\end{Shaded}

We omit the second question which is largely identical to the first, but
tests \(F(2)\) for the expected answer of \texttt{c(1,1)}.

The third file uses randomization to prevent students from hardcoding an
answer to \(F(n)\) for a given knowable value \(n\).

\begin{Shaded}
\begin{Highlighting}[]
\CommentTok{## @title Test F(n) for random n}
\CommentTok{## @score 2}

\KeywordTok{library}\NormalTok{(tinytest)}
\KeywordTok{using}\NormalTok{(ttdo)}

\NormalTok{n <-}\StringTok{ }\KeywordTok{sample}\NormalTok{(}\DecValTok{3}\OperatorTok{:}\DecValTok{20}\NormalTok{, }\DecValTok{1}\NormalTok{)}

\NormalTok{file <-}\StringTok{ "/grade/student/fib.R"}
\NormalTok{student <-}\StringTok{ }\NormalTok{plr}\OperatorTok{::}\KeywordTok{source_and_eval_safe}\NormalTok{(file,}
                                     \KeywordTok{fib}\NormalTok{(n), }\StringTok{"ag"}\NormalTok{)}

\KeywordTok{source}\NormalTok{(}\StringTok{"/grade/tests/ans.R"}\NormalTok{)}
\NormalTok{correct <-}\StringTok{ }\KeywordTok{fib}\NormalTok{(n)}

\KeywordTok{expect_equivalent_with_diff}\NormalTok{(student, correct)}
\end{Highlighting}
\end{Shaded}

It also shows another key feature: our use of the \textbf{diffobj}
package. We use a very small add-on package \textbf{ttdo} (an acronym
for `tinytest-diffobj') we created utilizing the extensions mechanism of
\textbf{tinytest} in order to provide more specific feedback in the test
results. The \textbf{ttdo} package is separate from our \textbf{plr}
package because of its potential use in contexts other than
PrairieLearn.

\begin{figure}
  \begin{center}
    \includegraphics[width=3in]{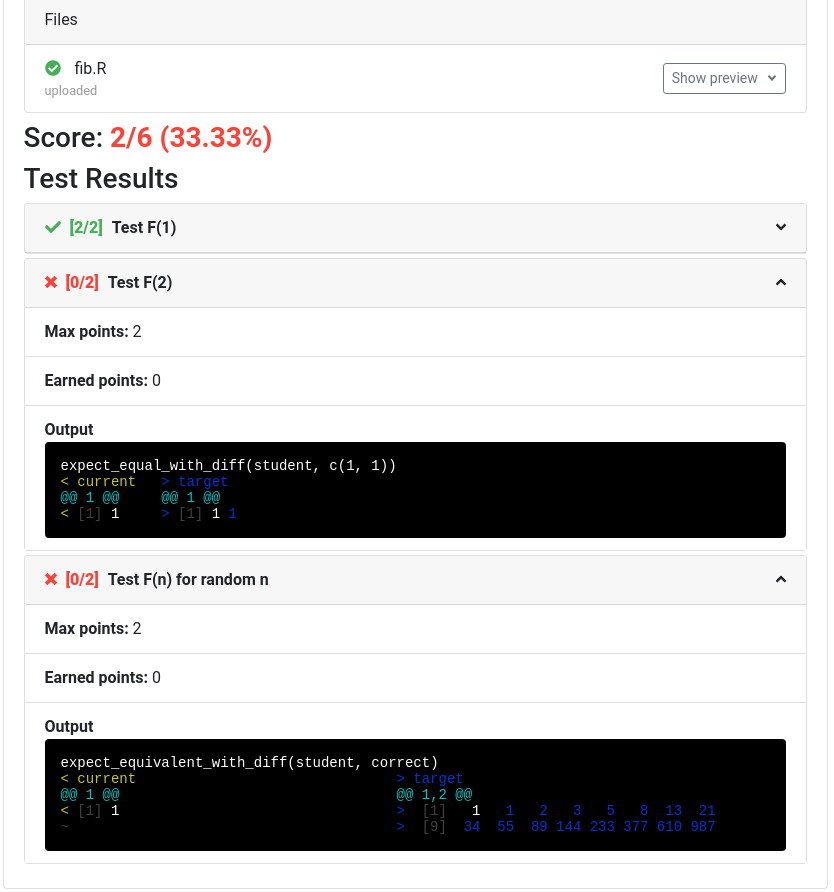} 
    \caption{Example Output of Autograder for Fibonacci Question}
    \label{fig:fib2ex} 
   \end{center}
\end{figure}

Figure \ref{fig:fib2ex} shows a screenshot resulting from providing an
answer that returns a content of 1 no matter the input. This passes
\(F(1)\), fails \(F(2)\) and fails \(F(n)\) for \(n>=3\). The screenshot
displays the effect of the colorized difference between the received
answer and the expected answer for the latter two questions.

\hypertarget{container}{%
\section{Container}\label{container}}

PrairieLearn allows for external containers. We use this feature to
deploy a custom container based on the \texttt{r-ubuntu} container from
the Rocker Project \citep{rocker:2017}. This container is setup with
access to the ``Personal Package Archive'' (PPA) by Michel Rutter which
provides a considerable subset of the R repositories (known as ``CRAN'')
as pre-compiled
binaries.\footnote{See the brief description at the top of
\url{https://CRAN.R-Project.org/bin/linux/ubuntu} for more details.}

Our \texttt{Dockerfile} is shown in Appendix 3. PrairieLearn always
checks for updated containers, so deployment of a new container is more
or less guaranteed. This also facilitates a workflow of incremental
changes as the `continuous deployment' is automated and relies on
trusted workflows supporting many other open source projects.

Similarly, by relying on widely-used and tested components such as the
Rocker Project containers as a base, along with provided Ubuntu
binaries, the risk of inadvertent breakage is minimized as well (when
compared to bespoke custom solutions not relying on more widely-used
compoents).

\hypertarget{security-aspects}{%
\section{Security Aspects}\label{security-aspects}}

R is a very flexible language that is somewhat difficult to sandbox as
it allows \emph{computation on the language}. Some approaches do
exist---the \textbf{RAppArmor} package \citep{CRAN:RAppArmor} wraps
around one of the two prevalent approaches for Linux is a candidate
given that we match the installation requirements by being on
Debian/Ubuntu systems.

Here, however, we opted for a more basic approach. \emph{All} files
copied in by \texttt{run.sh} are set to be owned by the \texttt{root}
user with no read, write or execution rights set for groups or others.
The one exception is the uploaded file containing the to-be-evaluated
student code. This file is then \texttt{source()}-ed in a lower-priority
process owned by the autograde user \texttt{ag}, and the supplied
function is evaluated with a given argument. We use the \textbf{unix}
package \citep{CRAN:unix} for this, taking advantage of the fact that
inside a container we are running as the superuser permitting us to
lower permissions. In other words, the one execution that could expose
secrets (of the untrusted code submitted by the student) is the one
running with the lowest possible permissions of the \texttt{ag} user
with all other files being ``locked-away'' and readable only by the
\texttt{root} user.

Concretely, our function \texttt{plr::source\_and\_eval\_safe()} shown
above relies on the function \texttt{unix::eval\_safe()} which takes
care of the (system-specific) details of process permission control. In
addition, we also minimize file permission changes. A sibbling function
\texttt{plr::eval\_safe\_as()} works similarly on an R expression rather
than file.

\hypertarget{summary}{%
\section{Summary}\label{summary}}

The PrairieLearn system \citep{talk:pl:2018} permits large-scale and
automated testing and grading of quizzes, exercises and tests as used in
university educated. It is designed as an open and extensible system.

We have created a custom autograding container for the R language to
both take advantage of the excellent PrairieLearn system, and extends
its facilities by using a unit testing framework which allows for
further customization. Our \textbf{plr} package \citep{R:plr} for R
autograding with PrairieLearn deployes the \textbf{tinytest} system
\citep{CRAN:tinytest} for unit testsing. It also extends it via the
\textbf{ttdo} package \citep{CRAN:ttdo} which permits the creation of
highly-informative \texttt{diff} objects produced by the
\textbf{diffobj} package \citep{CRAN:diffobj} which can be deployed
directly in the dataflow based on JSON objects used by PrairieLearn.

\bibliography{plr-intro}

\begin{thebibliography}{8}
\newcommand{\enquote}[1]{``#1''}
\providecommand{\natexlab}[1]{#1}
\providecommand{\url}[1]{\texttt{#1}}
\providecommand{\urlprefix}{URL }
\expandafter\ifx\csname urlstyle\endcsname\relax
  \providecommand{\doi}[1]{doi:\discretionary{}{}{}#1}\else
  \providecommand{\doi}{doi:\discretionary{}{}{}\begingroup
  \urlstyle{rm}\Url}\fi
\providecommand{\eprint}[2][]{\url{#2}}

\bibitem[{Boettiger and Eddelbuettel(2017)}]{rocker:2017}
Boettiger C, Eddelbuettel D (2017).
\newblock \enquote{{An Introduction to Rocker: Docker Containers for R}.}
\newblock \emph{{The R Journal}}, \textbf{9}(2), 527--536.
\newblock \doi{10.32614/RJ-2017-065}.
\newblock \urlprefix\url{https://doi.org/10.32614/RJ-2017-065}.

\bibitem[{Eddelbuettel and Barbehenn(2019{\natexlab{a}})}]{R:plr}
Eddelbuettel D, Barbehenn A (2019{\natexlab{a}}).
\newblock \emph{plr: Utility Functions for 'PrairieLearn' and R}.
\newblock R package version 0.0.2,
  \urlprefix\url{https://github.com/stat430dspm/plr}.

\bibitem[{Eddelbuettel and Barbehenn(2019{\natexlab{b}})}]{CRAN:ttdo}
Eddelbuettel D, Barbehenn A (2019{\natexlab{b}}).
\newblock \emph{ttdo: Extend 'tinytest' with 'diffobj'}.
\newblock R package version 0.0.4,
  \urlprefix\url{https://CRAN.R-project.org/package=ttdo}.

\bibitem[{Gaslam(2019)}]{CRAN:diffobj}
Gaslam B (2019).
\newblock \emph{diffobj: Diffs for R Objects}.
\newblock R package version 0.2.3,
  \urlprefix\url{https://CRAN.R-project.org/package=diffobj}.

\bibitem[{Ooms(2019{\natexlab{a}})}]{CRAN:RAppArmor}
Ooms J (2019{\natexlab{a}}).
\newblock \emph{RAppArmor: Bindings to AppArmor and Security Related Linux
  Tools}.
\newblock R package version 3.2,
  \urlprefix\url{https://CRAN.R-project.org/package=RAppArmor}.

\bibitem[{Ooms(2019{\natexlab{b}})}]{CRAN:unix}
Ooms J (2019{\natexlab{b}}).
\newblock \emph{unix: POSIX System Utilities}.
\newblock R package version 1.5,
  \urlprefix\url{https://CRAN.R-project.org/package=unix}.

\bibitem[{{van der Loo}(2019)}]{CRAN:tinytest}
{van der Loo} M (2019).
\newblock \emph{tinytest: Lightweight and Feature Complete Unit Testing
  Framework}.
\newblock R package version 1.0.0,
  \urlprefix\url{https://CRAN.R-project.org/package=tinytest}.

\bibitem[{Zilles \emph{et~al.}(2018)Zilles, West, Mussulman, and
  Bretl}]{talk:pl:2018}
Zilles C, West M, Mussulman D, Bretl T (2018).
\newblock \enquote{Making testing less trying: Lessons learned from operating a
  computer-based testing facility.}
\newblock In \emph{Proceedings of the 2018 Frontiers in Education Conference
  (FIE 2018)}.
\newblock
  \urlprefix\url{http://lagrange.mechse.illinois.edu/pubs/ZiWeMuBr2018/ZiWeMuBr2018.pdf}.

\end{thebibliography}
\bibliographystyle{jss}

\newpage
\onecolumn

\hypertarget{appendix-1-run.sh}{%
\subsection{\texorpdfstring{Appendix 1:
\texttt{run.sh}}{Appendix 1: run.sh}}\label{appendix-1-run.sh}}

\begin{Shaded}
\begin{Highlighting}[]
\CommentTok{#! /bin/bash}

\CommentTok{#### INIT}

\CommentTok{## the directory where the file pertaining to the job are mounted}
\VariableTok{JOB_DIR=}\StringTok{"/grade/"}
\CommentTok{## the other directories inside it}
\VariableTok{STUDENT_DIR=}\StringTok{"}\VariableTok{$\{JOB_DIR\}}\StringTok{student/"}
\VariableTok{AG_DIR=}\StringTok{"}\VariableTok{$\{JOB_DIR\}}\StringTok{serverFilesCourse/r_autograder/"}
\VariableTok{TEST_DIR=}\StringTok{"}\VariableTok{$\{JOB_DIR\}}\StringTok{tests/"}
\VariableTok{OUT_DIR=}\StringTok{"}\VariableTok{$\{JOB_DIR\}}\StringTok{results/"}

\CommentTok{## where we will copy everything}
\VariableTok{MERGE_DIR=}\StringTok{"}\VariableTok{$\{JOB_DIR\}}\StringTok{run/"}
\CommentTok{## where we will put the actual student code - this depends on what the autograder expects, etc}
\VariableTok{BIN_DIR=}\StringTok{"}\VariableTok{$\{MERGE_DIR\}}\StringTok{bin/"}

\CommentTok{## now set up the stuff so that our run.sh can work}
\BuiltInTok{echo} \StringTok{"[init] making directories"}
\FunctionTok{mkdir} \VariableTok{$\{MERGE_DIR\}} \VariableTok{$\{BIN_DIR\}} \VariableTok{$\{OUT_DIR\}}

\CommentTok{## making the test directory root:root and stripping group and others}
\CommentTok{## this will prevent the restricted user from snooping}
\FunctionTok{chown}\NormalTok{ -R root:root }\VariableTok{$\{TEST_DIR\}}
\FunctionTok{chmod}\NormalTok{ -R go-rwx    }\VariableTok{$\{TEST_DIR\}}

\CommentTok{## under 'tinytest' artefacts are created where the tests are running}
\CommentTok{## so let the 'ag' user own the directory to write files, run mkdir, ...}
\BuiltInTok{echo} \StringTok{"[init] setting up tests directory for 'ag' user"}
\FunctionTok{chown}\NormalTok{ ag:ag }\VariableTok{$\{TEST_DIR\}}\NormalTok{tests}

\BuiltInTok{echo} \StringTok{"[init] copying content"}
\FunctionTok{cp}    \VariableTok{$\{STUDENT_DIR\}}\NormalTok{* }\VariableTok{$\{BIN_DIR\}}
\FunctionTok{cp}    \VariableTok{$\{AG_DIR\}}\NormalTok{*      }\VariableTok{$\{MERGE_DIR\}}
\FunctionTok{cp}\NormalTok{ -r }\VariableTok{$\{TEST_DIR\}}\NormalTok{*    }\VariableTok{$\{MERGE_DIR\}}
\FunctionTok{chown}\NormalTok{ ag:ag           }\VariableTok{$\{MERGE_DIR\}}\NormalTok{tests}

\CommentTok{#### RUN}

\BuiltInTok{cd} \VariableTok{$\{MERGE_DIR\}}
\BuiltInTok{echo} \StringTok{"[run] starting autograder"}

\CommentTok{## we evaluate student code inside the test functions as a limited user called ag}
\CommentTok{## see the R package plr in the stat430dspm repo for details of the implementation}
\BuiltInTok{echo} \StringTok{"[run] Rscript pltest.R"}
\ExtensionTok{Rscript}\NormalTok{ pltest.R}

\KeywordTok{if}\BuiltInTok{ [} \OtherTok{!} \OtherTok{-s}\NormalTok{ results.json}\BuiltInTok{ ]}\NormalTok{; }\KeywordTok{then}
    \CommentTok{# Let's attempt to keep everything from dying completely}
    \BuiltInTok{echo}\NormalTok{ -n }\StringTok{'\{"succeeded": false, "score": 0.0, "message": "Catastrophic failure! '} \OperatorTok{>}\NormalTok{ results.json}
    \BuiltInTok{echo}    \StringTok{'Contact course staff and have them check the logs for this submission."\}'} \OperatorTok{>>}\NormalTok{ results.json}
\KeywordTok{fi}

\BuiltInTok{echo} \StringTok{"[run] autograder completed"}

\CommentTok{# get the results from the file}
\FunctionTok{cp} \VariableTok{$\{MERGE_DIR\}}\NormalTok{/results.json }\StringTok{'/grade/results/results.json'}
\BuiltInTok{echo} \StringTok{"[run] copied results"}
\end{Highlighting}
\end{Shaded}

\newpage

\hypertarget{appendix-2-pltest.r}{%
\subsection{\texorpdfstring{Appendix 2:
\texttt{pltest.R}}{Appendix 2: pltest.R}}\label{appendix-2-pltest.r}}

\begin{Shaded}
\begin{Highlighting}[]
\CommentTok{## Simple-yet-good enough runner for R questions}
\CommentTok{##}
\CommentTok{## Alton Barbehenn and Dirk Eddelbuettel, Aug/Sep 2019}

\NormalTok{message_to_test_result <-}\StringTok{ }\ControlFlowTok{function}\NormalTok{(msg, }\DataTypeTok{mxpts=}\DecValTok{100}\NormalTok{) \{}
    \KeywordTok{data.frame}\NormalTok{(}
        \DataTypeTok{name =} \StringTok{"Error"}\NormalTok{,}
        \DataTypeTok{max_points =}\NormalTok{ mxpts,}
        \DataTypeTok{points =} \DecValTok{0}\NormalTok{,}
        \DataTypeTok{output =}\NormalTok{ msg}\OperatorTok{$}\NormalTok{message}
\NormalTok{    )}
\NormalTok{\}}


\NormalTok{result <-}\StringTok{ }\KeywordTok{tryCatch}\NormalTok{(\{}

    \CommentTok{## Set seed for control over randomness (change every day)}
    \KeywordTok{set.seed}\NormalTok{(}\KeywordTok{as.integer}\NormalTok{(}\KeywordTok{Sys.Date}\NormalTok{()))}

    \CommentTok{## Directory with test files}
\NormalTok{    tests_dir <-}\StringTok{ "/grade/tests/tests"}

    \CommentTok{## Get question information on available points and displayed title}
\NormalTok{    question_details <-}\StringTok{ }\NormalTok{plr}\OperatorTok{::}\KeywordTok{get_question_details}\NormalTok{(tests_dir)}

    \CommentTok{## Run tests in the test directory}
    \KeywordTok{cat}\NormalTok{(}\StringTok{"[pltest] about to call tests from"}\NormalTok{, }\KeywordTok{getwd}\NormalTok{(), }\StringTok{"}\CharTok{\textbackslash{}n}\StringTok{"}\NormalTok{)}
\NormalTok{    test_results <-}\StringTok{ }\KeywordTok{as.data.frame}\NormalTok{(tinytest}\OperatorTok{::}\KeywordTok{run_test_dir}\NormalTok{(tests_dir, }\DataTypeTok{verbose =} \OtherTok{FALSE}\NormalTok{))}

    \CommentTok{## Aggregate test results and process NAs as some question may have exited}
\NormalTok{    res <-}\StringTok{ }\KeywordTok{merge}\NormalTok{(test_results, question_details, }\DataTypeTok{by =} \StringTok{"file"}\NormalTok{, }\DataTypeTok{all =} \OtherTok{TRUE}\NormalTok{)}
    \CommentTok{## Correct answers get full points, other get nothing}
\NormalTok{    res}\OperatorTok{$}\NormalTok{points <-}\StringTok{ }\KeywordTok{ifelse}\NormalTok{( }\OperatorTok{!}\KeywordTok{is.na}\NormalTok{(res}\OperatorTok{$}\NormalTok{result) }\OperatorTok{&}\StringTok{ }\NormalTok{res}\OperatorTok{$}\NormalTok{result}\OperatorTok{==}\OtherTok{TRUE}\NormalTok{,  res}\OperatorTok{$}\NormalTok{max_points, }\DecValTok{0}\NormalTok{)}
    \CommentTok{## For false answers we collate call and diff output (from diffobj::diffPrint)}
\NormalTok{    res}\OperatorTok{$}\NormalTok{output <-}\StringTok{ }\KeywordTok{ifelse}\NormalTok{( }\OperatorTok{!}\KeywordTok{is.na}\NormalTok{(res}\OperatorTok{$}\NormalTok{result) }\OperatorTok{&}\StringTok{ }\NormalTok{res}\OperatorTok{$}\NormalTok{result}\OperatorTok{==}\OtherTok{FALSE}\NormalTok{,}
                         \KeywordTok{paste}\NormalTok{(res}\OperatorTok{$}\NormalTok{call, res}\OperatorTok{$}\NormalTok{diff, }\DataTypeTok{sep =} \StringTok{"}\CharTok{\textbackslash{}n}\StringTok{"}\NormalTok{), }\StringTok{""}\NormalTok{)}
\NormalTok{    score <-}\StringTok{ }\KeywordTok{sum}\NormalTok{(res}\OperatorTok{$}\NormalTok{points) }\OperatorTok{/}\StringTok{ }\KeywordTok{sum}\NormalTok{(res}\OperatorTok{$}\NormalTok{max_points) }\CommentTok{# total score}

    \CommentTok{## Columns needed by PL}
\NormalTok{    res <-}\StringTok{ }\NormalTok{res[, }\KeywordTok{c}\NormalTok{(}\StringTok{"name"}\NormalTok{, }\StringTok{"max_points"}\NormalTok{, }\StringTok{"points"}\NormalTok{, }\StringTok{"output"}\NormalTok{)]}

    \CommentTok{## output}
    \KeywordTok{list}\NormalTok{(}\DataTypeTok{tests =}\NormalTok{ res, }\DataTypeTok{score =}\NormalTok{ score, }\DataTypeTok{succeeded =} \OtherTok{TRUE}\NormalTok{)}
\NormalTok{\},}
\DataTypeTok{warning =} \ControlFlowTok{function}\NormalTok{(w) }\KeywordTok{list}\NormalTok{(}\DataTypeTok{tests =} \KeywordTok{message_to_test_result}\NormalTok{(w), }\DataTypeTok{score =} \DecValTok{0}\NormalTok{, }\DataTypeTok{succeeded =} \OtherTok{FALSE}\NormalTok{),}
\DataTypeTok{error =} \ControlFlowTok{function}\NormalTok{(e) }\KeywordTok{list}\NormalTok{(}\DataTypeTok{tests =} \KeywordTok{message_to_test_result}\NormalTok{(e), }\DataTypeTok{score =} \DecValTok{0}\NormalTok{, }\DataTypeTok{succeeded =} \OtherTok{FALSE}\NormalTok{) )}

\CommentTok{## Record results as the required JSON object}
\NormalTok{jsonlite}\OperatorTok{::}\KeywordTok{write_json}\NormalTok{(result, }\DataTypeTok{path =} \StringTok{"results.json"}\NormalTok{, }\DataTypeTok{auto_unbox =} \OtherTok{TRUE}\NormalTok{, }\DataTypeTok{force =} \OtherTok{TRUE}\NormalTok{)}
\end{Highlighting}
\end{Shaded}

\newpage

\hypertarget{appendix-3-dockerfile-for-stat430pl-container}{%
\subsection{\texorpdfstring{Appendix 3: \texttt{Dockerfile} for
stat430/pl
container}{Appendix 3: Dockerfile for stat430/pl container}}\label{appendix-3-dockerfile-for-stat430pl-container}}

\begin{Shaded}
\begin{Highlighting}[]
\CommentTok{# Image used for PrairieLearn external grading of R questions}
\CommentTok{# as well as general support of STAT 430 Data Science Programming Methods}

\CommentTok{# Alton Barbehenn and Dirk Eddelbuettel, 2019}

\CommentTok{# Before we based our image on prairielearn/centos7-python, }
\CommentTok{# and that worked, but it was harder to maintian and a lot }
\CommentTok{# than we needed. Now we're using rocker/tidyverse as our}
\CommentTok{# base because it's more focused and solves the prerequisites}
\CommentTok{# for us, along with providing many useful R packages. }

\ExtensionTok{FROM}\NormalTok{ rocker/r-ubuntu:18.04}

\CommentTok{# From prairielearn/centos7-python: Needed to properly handle UTF-8}
\ExtensionTok{ENV}\NormalTok{ PYTHONIOENCODING=UTF-8}


\CommentTok{# Install required libraries -- using prebuild binaries where available}
\ExtensionTok{RUN}\NormalTok{ apt-get update }\KeywordTok{&&} \ExtensionTok{apt-get}\NormalTok{ install -y \textbackslash{}}
\NormalTok{    git \textbackslash{}}
\NormalTok{    r-cran-data.table \textbackslash{}}
\NormalTok{    r-cran-devtools \textbackslash{}}
\NormalTok{    r-cran-doparallel \textbackslash{}}
\NormalTok{    r-cran-dygraphs \textbackslash{}}
\NormalTok{    r-cran-foreach \textbackslash{}}
\NormalTok{    r-cran-fs \textbackslash{}}
\NormalTok{    r-cran-future.apply \textbackslash{}}
\NormalTok{    r-cran-gh \textbackslash{}}
\NormalTok{    r-cran-git2r \textbackslash{}}
\NormalTok{    r-cran-igraph \textbackslash{}}
\NormalTok{    r-cran-memoise \textbackslash{}}
\NormalTok{    r-cran-microbenchmark \textbackslash{}}
\NormalTok{    r-cran-png \textbackslash{}}
\NormalTok{    r-cran-rcpparmadillo \textbackslash{}}
\NormalTok{    r-cran-rex \textbackslash{}}
\NormalTok{    r-cran-rsqlite \textbackslash{}}
\NormalTok{    r-cran-runit \textbackslash{}}
\NormalTok{    r-cran-shiny \textbackslash{}}
\NormalTok{    r-cran-stringdist \textbackslash{}}
\NormalTok{    r-cran-testthat \textbackslash{}}
\NormalTok{    r-cran-tidyverse \textbackslash{}}
\NormalTok{    r-cran-tinytest \textbackslash{}}
\NormalTok{    r-cran-xts \textbackslash{}}
\NormalTok{    sqlite3 \textbackslash{}}
\NormalTok{    sudo}

\CommentTok{# Install additional R packages from CRAN (on top of the ones pre-built as r-cran-*)}
\ExtensionTok{RUN}\NormalTok{ install.r bench diffobj flexdashboard lintr ttdo unix}

\CommentTok{# Install plr -- for now (?) from GH; also install visualTest}
\ExtensionTok{RUN}\NormalTok{ installGithub.r stat430dspm/plr MangoTheCat/visualTest}

\ExtensionTok{RUN}\NormalTok{ useradd ag }\DataTypeTok{\textbackslash{} }
    \KeywordTok{&&} \FunctionTok{mkdir}\NormalTok{ /home/ag \textbackslash{}}
    \KeywordTok{&&} \FunctionTok{chown}\NormalTok{ ag:ag /home/ag \textbackslash{}}
    \KeywordTok{&&} \BuiltInTok{echo} \StringTok{"[user]"} \OperatorTok{>}\NormalTok{ /home/ag/.gitconfig \textbackslash{}}
    \KeywordTok{&&} \BuiltInTok{echo} \StringTok{"   name = Autograding User"} \OperatorTok{>>}\NormalTok{ /home/ag/.gitconfig \textbackslash{}}
    \KeywordTok{&&} \BuiltInTok{echo} \StringTok{"   email = ag@nowhere"} \OperatorTok{>>}\NormalTok{ /home/ag/.gitconfig \textbackslash{}}
    \KeywordTok{&&} \FunctionTok{chown}\NormalTok{ ag:ag /home/ag/.gitconfig}
    
\end{Highlighting}
\end{Shaded}

%\showmatmethods

\end{document}